\documentclass[%
 reprint,
superscriptaddress,
longbibliography,
nofootinbib,
 amsmath,amssymb,
 aps,
 showkeys,
]{revtex4-1}

\usepackage{graphicx}% Include figure files
\usepackage{dcolumn}% Align table columns on decimal point
\usepackage{bm}% bold math
\usepackage{mhchem}
\usepackage{color}
\usepackage{siunitx}
\usepackage{mathtools}
\usepackage{upgreek} % to get upright mu in micron units (see below)
\usepackage{lineno}
\usepackage{xfrac}
\usepackage{commath}
\usepackage[normalem]{ulem}

\begin{document}

\preprint{APS/123-QED}

\title{Shape- and orientation-dependent diffusiophoresis of colloidal ellipsoids}

\author{Viet Sang Doan}%
\affiliation{Department of Mechanical and Aerospace Engineering, University at Buffalo, The State University of New York, Buffalo, New York 14260, USA}

\author{Dong-Ook Kim}%
\affiliation{Department of Mechanical Engineering and Mechanics, Drexel University, Philadelphia, Pennsylvania 19104, USA}

\author{Craig Snoeyink}%
\affiliation{Department of Mechanical and Aerospace Engineering, University at Buffalo, The State University of New York, Buffalo, New York 14260, USA}

\author{Ying Sun}%
\affiliation{Department of Mechanical Engineering and Mechanics, Drexel University, Philadelphia, Pennsylvania 19104, USA}

\author{Sangwoo Shin}%
\email{\textbf{E-mail:} sangwoos@buffalo.edu}
\affiliation{Department of Mechanical and Aerospace Engineering, University at Buffalo, The State University of New York, Buffalo, New York 14260, USA}

%\date{\today}% It is always \today, today,
             %  but any date may be explicitly specified
%\keywords{TBD}

\begin{abstract} 
We present the diffusiophoresis of ellipsoidal particles induced by ionic solute gradients. 
Contrary to the common expectation that diffusiophoresis is shape independent, here we show experimentally that this assumption breaks down when the thin Debye layer approximation is relaxed. 
By tracking the translation and rotation of various ellipsoids, we find that the phoretic mobility of ellipsoids is sensitive to the eccentricity and the orientation of the ellipsoid relative to the imposed solute gradient, and can further lead to nonmonotonic behavior under strong confinement. We show that such a shape- and orientation-dependent diffusiophoresis of colloidal ellipsoids can be easily captured  by modifying theories for spheres. 
\end{abstract}

\maketitle 
Phoretic propulsion refers to a directional motion of a colloidal particle set by the fluid flow at the particle-liquid interface due to the nonequilibrium thermodynamic forces \cite{Anderson1989}. Among ways to drive the interfacial flow, such as via electric stress (electrophoresis) and surface tension gradients (Marangoni propulsion), osmotic pressure imbalance induced by solute concentration gradients can also drive interfacial flow along the particle surface, thereby causing freely suspended particles to migrate. 
This process, known as diffusiophoresis \cite{Derjaguin1961}, has been recently recognized in biological settings for controlling the motion of biomolecules and biocolloids, such as proteins \cite{annunziata2012protein,lechlitner2018macromolecule,ramm2021diffusiophoretic,peter2022microscale}, nucleic acids \cite{Palacci2010,Friedrich2017,Shin2017c}, liposomes \cite{Shin2019,Rasmussen2020}, bacteria \cite{Lee2018,Doan2020}, and blood cells \cite{Hartman2018},
due to the abundance of chemicals and their gradients in biological systems \cite{keenan2008biomolecular,Sear2019}.

While many biocolloids and other naturally existing particles (e.g., clay) are often non-spherical, 
it is a common practice to treat them as a sphere when analyzing their phoretic motion, although earlier theoretical studies have pointed out that the diffusiophoresis of nonspherical particles may differ from spheres under a range of particle and solute conditions \cite{dukhin1993non,keh1993diffusiophoresis,hsu2010diffusiophoresis,keh1993diffusiophoresiscylinder,joo2010diffusiophoresis,hsu2012importance,popescu2010phoretic}. 
In this Letter, we investigate experimentally the diffusiophoretic transport of microscale colloidal ellipsoids. Using single-particle tracking in microfluidic settings, we find that the diffusiophoresis of ellipsoids is not only different from their spherical counterpart, but also shape- and orientation-dependent, and can further lead to non-monotonic mobility under strong confinement. 
 
Colloidal ellipsoids were obtained by uniaxially stretching spherical polystyrene beads into prolate ellipsoids \cite{ho1993preparation,kim2016deposition}.
In brief, carboxylate-modified polystyrene beads (radius $a$ = 0.5 $\mu$m) were embedded in a thin film of polyvinyl alcohol (PVA), which were subsequently stretched to the desired strain level at a temperature above the glass transition point. 
Upon cooling, the elongated particles were recovered by dissolving the PVA film, followed by reoxidizing the particles to restore the native surface charge (particle zeta potential $\zeta\approx-60$ mV) \cite{ho1997examination}.
Prolate ellipsoids of aspect ratios ($=a_1/a_2$, where $a_1$ and $a_2$ are, respectively, semi-major and semi-minor axes) of 3.5 and 6.5 were fabricated (Fig.~1a). Due to the nature of the fabrication method, these ellipsoids have the same volume.

\begin{figure}[t!]
	\centering
	\includegraphics[width=8.5cm]{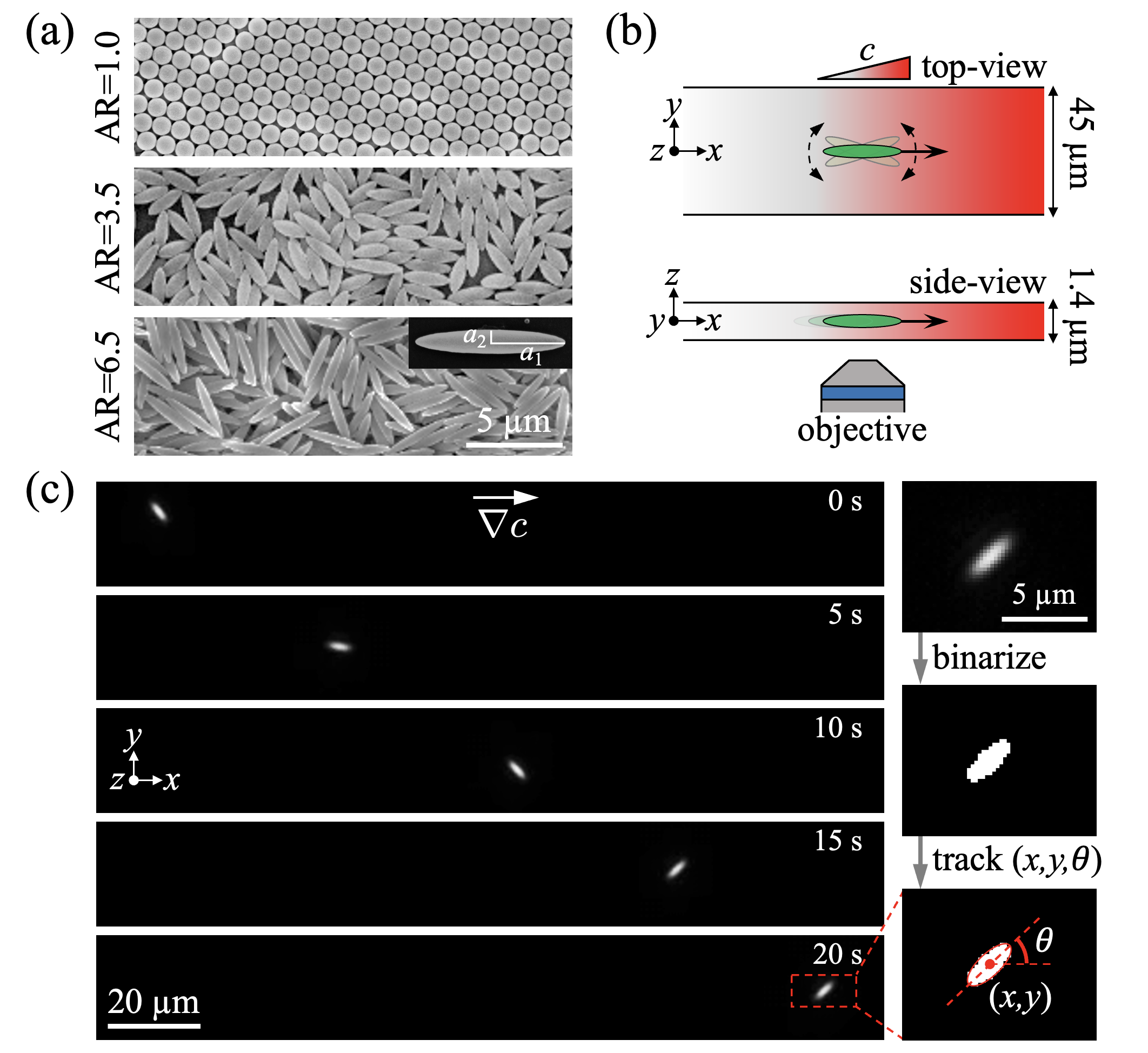} %update
	\caption{
		\textbf{Experimental setup for characterizing the translation and rotation of colloidal ellipsoid under solute gradients.}	
		(a) Scanning electron microscope (SEM) images of prolate ellipsoidal particles of equal volume used in this study. The particle aspect ratios (AR~=~$a_1/a_2$) are 1.0 (sphere), 3.5, and 6.5.
		(b) A Hele-Shaw-type microfluidic channel (width = 45~$\mu$m, height = 1.4~$\mu$m) that is confined in the $z$ direction is used to perform diffusiophoresis experiments. The shallow height of the channel effectively constrains the ellipsoid rotation about the $y$ axis. 
		(c) An image sequence showing an ellipsoidal particle of AR~=~6.5 undergoing diffusiophoresis in the positive $x$ direction via LiCl concentration gradients. To analyze the particle position and orientation, each image is binarized to identify the particle boundary, and then track its centroid $(x,y)$ and the orientation angle of the major axis with respect to the $x$ axis ($\theta$). 
	}
	\label{fig:mobility}
\end{figure}

To induce diffusiophoresis, we form a one-dimensional (1D) solute gradient in a Hele-Shaw-like microfluidic channel, where the width and height of the channel are, respectively, 45~$\mu$m and 1.4~$\mu$m (Fig. 1b). 
The channel is initially filled with a LiCl solution ($c_i=1$~mM) containing colloidal particles, which is followed by introducing a LiCl solution of lower concentration ($c_o=0.1$~mM) to create the solute concentration gradient along the channel (in the positive $x$-direction), thus inducing particle diffusiophoresis. LiCl was chosen to induce strong electrolytic diffusiophoresis \cite{Abecassis2008,Doan2020}.
The entire channel surfaces were coated with a thin layer of PVA to prevent the particles from adhering to the channel wall \cite{trantidou2017hydrophilic,park2021microfluidic}. This also effectively suppresses any undesired diffusioosmotic flow occurring along the channel wall, which influences the particle transport \cite{Shin2017b,Shin2020a,alessio2021diffusiophoresis,migacz2022diffusiophoresis}. 

Due to the shallow channel height, the ellipsoids are confined in the height direction ($z$-direction) such that their motion is effectively two dimensional (2D).
This condition makes it possible to track the in-plane translation and rotation of individual ellipsoids with a wide-field microscope. 
A typical experiment is shown in Fig.~1c, where an ellipsoid of AR~=~6.5 is migrating along the LiCl  gradients imposed in the $x$ direction. By binarizing the microscopy images, we can track the instantaneous particle position $[x(t),y(t)]$ and the angle of the major axis with respect to the $x$ axis, $\theta(t)$.

An intriguing observation was made where we found a strong correlation between the $x$-direction velocity $v_x$ and the orientation angle $\theta$ of the ellipsoid during its diffusiophoretic migration (Figs.~2a,b). 
As the ellipsoid's major axis aligns  with the $x$ axis ($\theta \rightarrow 0$), the particle velocity tends to increase, whereas the particle velocity decreases as the ellipsoid tilts away from the $x$ axis ($|\theta| \rightarrow 90^\circ$). 
We analyze the particle motion frame by frame to extract the instantaneous diffusiophoretic mobility $M_i(t)$, which is defined as $M_i(t)=v_x(t)\cdot c(x,t)/\partial_x c(x,t)$, where the local solute concentration $c$ and its gradient $\partial_xc$ are obtained analytically \cite{Doan2020}.
 As we plot the instantaneous diffusiophoretic mobility $M_i$ against $\theta$ shown in Fig.~2c, this orientation dependence becomes more evident.

\begin{figure}[t!]
	\centering
	\includegraphics[width=8.5cm]{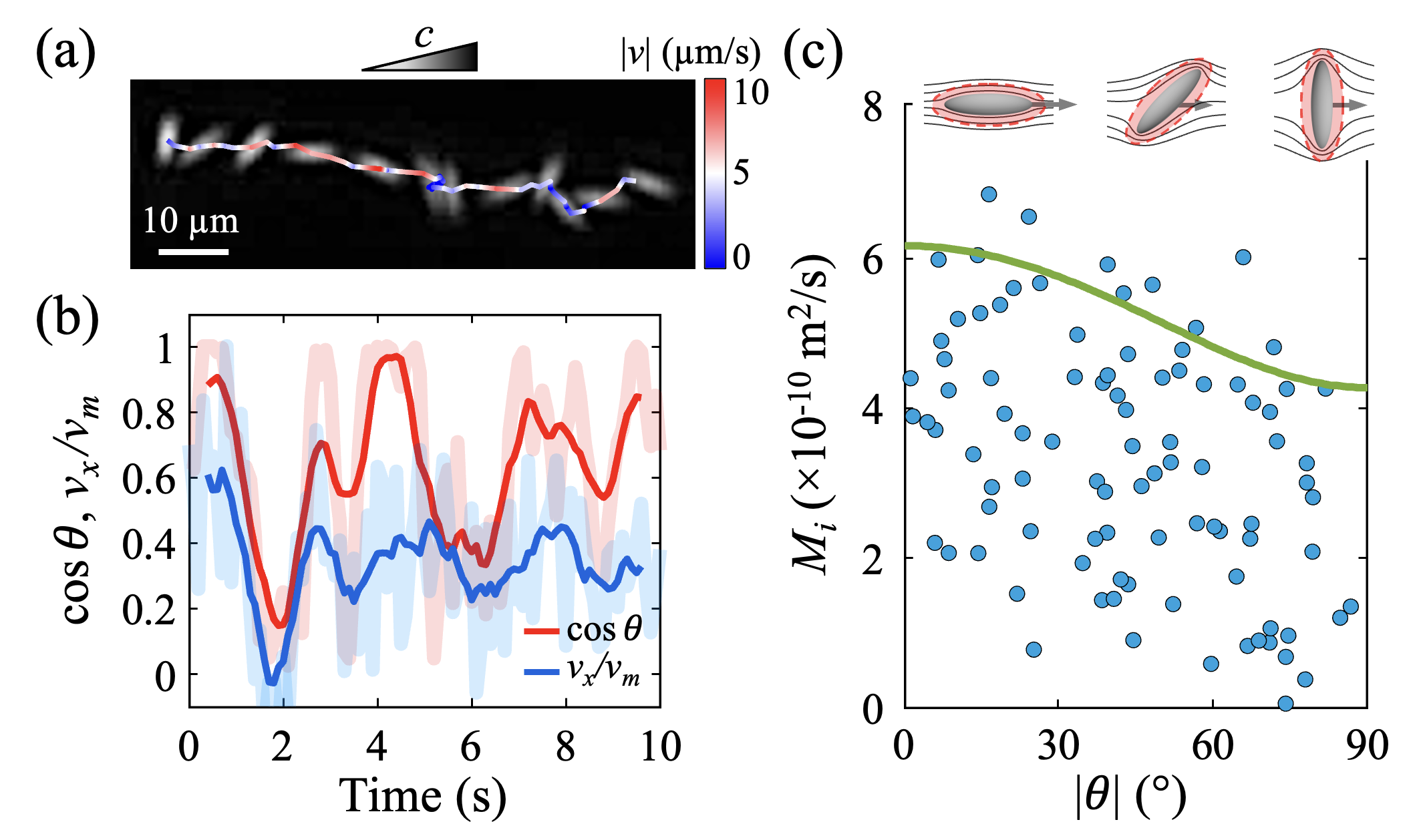} 
	\caption{
		\textbf{Orientation-dependent diffusiophoresis of ellipsoidal particles.}	
		(a) A time-lapse image (10~s interval between successive images) of an ellipsoid (AR~=~6.5) undergoing diffusiophoresis.  The particle trajectory is overlaid on the image. Color represents velocity magnitude.
		(b) A plot of particle velocity in the $x$ direction $v_x$ (normalized by maximum velocity $v_m$) and $\mathrm{cos}\,\theta$ showing a strong correlation between the two. 
		(c) The instantaneous diffusiophoretic mobility $M_i$ plotted with respect to $|\theta|$ at that instance extracted from a single trajectory. The green curve is Eq.~(6). The inset shows flux lines computed numerically by solving the Laplace equation around each ellipsoid. The red layer around the ellipsoid indicates the Debye layer of constant thickness.
	}
	\label{fig:mobility2}
\end{figure}

This experimental observation of orientation-dependent diffusiophoresis of non-spherical colloids is partially in accordance with earlier theoretical studies on electrophoresis of slender particles, including cylinders and prolate ellipsoids in which a particle is expected to migrate faster along the symmetry axis (major axis aligned with the imposed electric field) than transverse to the axis provided the thin Debye layer assumption is relaxed \cite{henry1931cataphoresis,yoon1989electrophoresis,dukhin1993non}. 
Similar predictions have been made  for diffusiophoresis by Keh and others, which also showed numerically that the longitudinal mobility (along the major axis) of prolate ellipsoids \cite{keh1993diffusiophoresis,hsu2010diffusiophoresis} and cylinders \cite{keh1993diffusiophoresiscylinder,joo2010diffusiophoresis,hsu2012importance} is faster than the transverse mobility (along the minor axis).

Such anisotropic behavior is primarily attributed to how the solute transport and the associated interfacial flow within the Debye layer changes depending on the orientation angle, which directly impacts the driving force for diffusiophoresis. 
For instance, the insets in Fig.~2c illustrate that for an otherwise 1D gradient, the solute flux lines residing within the Debye layer (red layer) mostly conform to the ellipsoid when laid parallel to the field gradient, making most of the gradients (and thus the driving force) tangential to the ellipsoid surface. 
In contrast, as the ellipsoid tilts away from the gradient, the flux lines gradually become less tangential to the particle surface, thereby weakening the phoretic migration.

Despite the translational phoretic motion of ellipsoids being sensitive to the particle orientation, the probability of occurrence of a particular orientation angle $P(\theta)$ appears to be more or less uniform, indicating no preferential orientation while experiencing diffusiophoresis. This is shown in Fig.~3a as a histogram of angle distribution during diffusiophoresis (10$\times$ solute concentration difference; right panel) and pure diffusion (no concentration difference; left panel). The lack of preferential orientation angle implies a negligible effect of diffusiophoresis on the particle rotation. While this is an anticipated behavior for a uniformly charged ellipsoid in unidirectional field gradients \cite{obrien1988electrophoresis,solomentsev1994electrophoresis}, it is not necessarily the case for diffusiophoresis even under a 1D linear gradient due to the logarithmic nature of diffusiophoresis \cite{Palacci2012,Shin2018}.

\begin{figure}[t!]
	\centering
	\includegraphics[width=8.5cm]{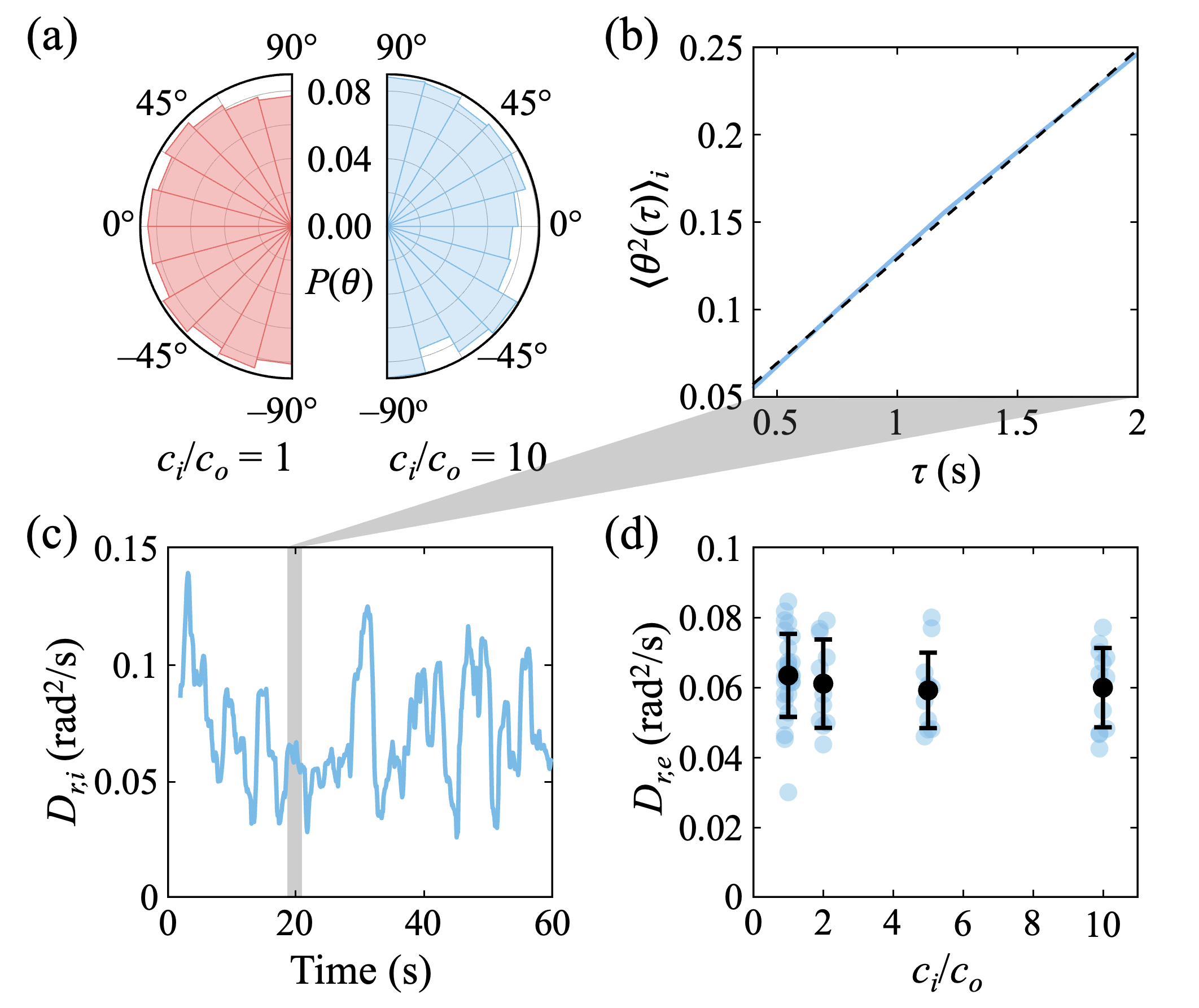}
	\caption{
		\textbf{Diffusiophoresis  has negligible influence on the particle rotation.}	
		(a) Histogram of probability of orientation angle $P(\theta)$ for diffusion (red) and diffusiophoresis (blue; 10$\times$ solute concentration difference, $c_i/c_o = 10$).
		(b) The instantaneous rotational diffusivity $D_{r,i}$ is obtained by measuring the mean-square angular displacement with a lag time of up to 2 s. The dashed black line is a linear fit to calculate $D_{r,i}$. 
		(c) Instantaneous rotational diffusivity $D_{r,i}$ over time obtained from a single-particle trajectory. 
		(d) Time-averaged effective rotational diffusivity of ellipsoidal particles $D_{r,e}$ with a range of solute concentration contrast $c_i/c_o$. The case in which there is no concentration difference ($c_i/c_o=1$) represents  pure diffusion. 
	}
	\label{fig:mobility2}
\end{figure}

We evaluate the effect of diffusiophoresis on the particle rotation by measuring the rotational diffusivity of the ellipsoid during its translation via either diffusiophoresis or diffusion.
We obtain the instantaneous rotational diffusivity $D_{r,i}$ of the ellipsoids by measuring the instantaneous mean-square angular displacement $\langle \theta^2\rangle_i$ with the running lag time $\tau$ of up to 2~s (Fig.~3b), where $D_{r,i}=\langle \theta^2\rangle_i /2\tau$. As shown in Fig.~3c, we do not observe any noticeable change in the overall tendency of the rotational diffusivity over time.

By time-averaging $D_{r,i}$ over the course of observation (which typically lasts for several minutes), we obtain the effective rotational diffusivity $D_{r,e}=\int D_{r,i} \,dt/ \int dt$ across a range of solute concentration contrast $c_i/c_o$. Figure~3d shows that the effective rotational diffusivity remains invariant regardless of the solute contrast, confirming the negligible influence of diffusiophoresis on the rotational diffusion.
This also makes the ellipsoids migrate predominantly in a unidirectional manner along the gradient. Despite the anisotropic mobility, which generally causes a particle to migrate at an angle with respect to the driving force in viscous environments \cite{happel2012low}, the negligible influence of diffusiophoresis on the particle rotation makes the particle to migrate straight along the gradient as the Brownian fluctuation randomizes any preferential orientation.
For instance, the particle rotation due to diffusiophoresis must arise from an asymmetry in the diffusioosmotic flow around the ellipsoid, possibly by the logarithmic dependence of diffusioosmosis. This velocity variation around the ellipsoid due to such behavior is estimated as $dv\approx \frac{M}{a} (\frac{dc}{c})^2\sim 0.01$~$\mu$m/s, which gives the rotational Peclet number as $\mathrm{Pe}_r =\frac{dv/a}{D_r} = \mathcal{O}(0.1)$, confirming that the rotational dynamics of the colloidal ellipsoids is mainly governed by the Brownian fluctuation.

Moving on, we measure the diffusiophoresis of colloidal ellipsoids of varying aspect ratios (AR~=~1.0 (sphere), 3.5, and 6.5), as shown in Fig. 4a. 
From over 10,000 $M_i$ of various ellipsoids, we extract the ``principal'' mobility along the longitudinal ($\parallel$) and transverse ($\perp$) direction, which are the mobility for which $\abs{\theta}<15^\circ$ and $\abs{\theta}>75^\circ$, respectively.
As expected, the longitudinal mobility is larger than the transverse mobility, i.e., $M_{i}^\parallel>M_{i}^\perp$, for both ellipsoids (AR~=~3.5 and AR~=~6.5). 
The mobility contrast becomes more significant with increasing AR, where the difference between 
$M^\parallel$ and $M^\perp$ for ellipsoids of AR~=~6.5 is larger than for AR~=~3.5. 

\begin{figure}[t!]
	\centering
	\includegraphics[width=8.5cm]{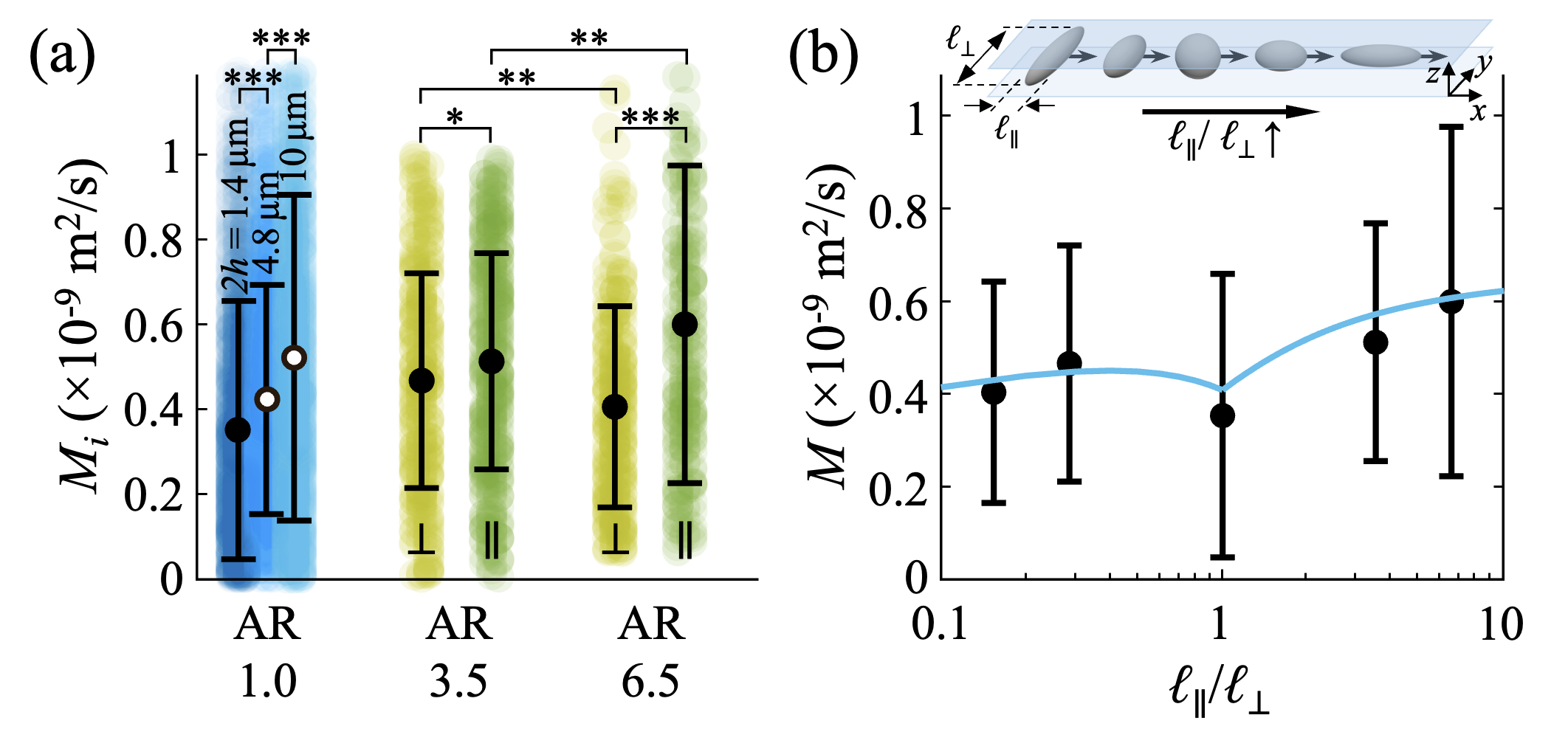}
	\caption{
		\textbf{Shape-dependent diffusiophoretic mobility of ellipsoidal particles.}	
		(a) Instantaneous mobility $M_i$ for particles of various aspect ratios. 
		Black circles and the corresponding error bars represent the mean and standard deviation.
		Symbols $\parallel$ and $\perp$ represent the mobility along the major and minor axis, respectively.
		Solid circles indicate experiments conducted in $2h=1.4$~$\mu$m channel. Open circles are sphere experiments conducted in $2h=4.8$~$\mu$m and $10$~$\mu$m channels. 
		$^{***}p < 0.001$, $^{**}p < 0.01$, $^{*}p < 0.05$, unpaired two-tailed t-test.
		(b) Diffusiophoretic mobility with respect to $\ell^{\parallel}/\ell^{\perp}$, where $\ell^{\parallel}$ and $\ell^{\perp}$ are the length of the ellipsoid parallel and perpendicular to the migration direction, respectively.
		The blue curve represents Eq.~(1).
		}
	\label{fig:mobility}
\end{figure}

Notably, our data show that the mobility of ellipsoidal particles, both the longitudinal  and transverse mobilities, is larger than the spherical counterpart. 
This is clearly shown in Fig.~4b, where the average mobility data from Fig.~4a are replotted with respect to the ratio of the principal axis along the gradient direction $\ell^{\parallel}$  to the principal axis transverse to the gradient $\ell^{\perp}$, i.e., $\ell^{\parallel}/\ell^{\perp}$.
The observed nonmonotonic tendency is in direct contrast to earlier theoretical studies, many of which predict that the sphere mobility is larger than the transverse mobility, but smaller than the longitudinal mobility under constant volume conditions \cite{harris1970simplified,yoon1989electrophoresis,keh1993diffusiophoresis,ohshima2021approximate}. 
Given that previous theoretical studies consider unbounded transport, we speculate that this discrepancy is due to the physical confinement imposed by the shallow channel in our system. The rationale is that the ellipsoids will experience less boundary confinement from the walls compared to their spherical counterpart having the same volume, as the semi-minor axis of a prolate ellipsoid is smaller than the original sphere radius of same volume, i.e., $a_2<a$, such that the average gap spacing between the particle surface and the channel wall for an ellipsoid is larger than that for a sphere (see the inset in Fig.~4b, for example).
We verify the existence of the confinement effect in our system by measuring the diffusiophoretic mobility of spheres ($2a=1$~$\mu$m) in channels of different heights, ranging from $2h = 1.4$~$\mu$m to 10~$\mu$m. 
Shown as the blue data in Fig.~4a, there is a considerable increase in the sphere mobility with increasing channel height, directly confirming the confinement effect on diffusiophoresis. 

Together, these experimental observations indicate that the particle motion in our system is influenced by the particle shape and the degree of confinement. These two effects are often strongly coupled when the Debye layer is relatively thick because the boundary deforms the Debye layer, thus making it formidable to solve the coupled, nonlinear transport equations analytically \cite{doan2021confinement}. 
However, the coupled dynamics may be relaxed when the Debye layer is thin relative to the local radius of curvature of the particle surface and the particle is weakly charged such that $\mathrm{exp}(ze|\zeta|/2kT)/\kappa a \approx 0.08 \ll 1$, where $z$ is the valence, $e$ is the elementary charge, $k$ is the Boltzmann constant, and $T$ is the temperature \cite{dukhin1993non}.
From this we make an assumption that the mobility can be expressed as a separable form,
\begin{equation}
	M(\lambda_1,\lambda_2)\approx M_\infty  \Phi_1(\lambda_1) \Phi_2(\lambda_2),
\end{equation}
where $M_\infty$ is the diffusiophoretic mobility of an unbounded ellipsoid in the limit of infinitesimally thin Debye layer, $\Phi_1$ is the dimensionless factor that accounts for the finite Debye layer effect (curvature effect) in the absence of confinement, and $\Phi_2$ is the dimensionless factor accounting for the confinement effect in the absence of the curvature effect in the limit of infinitesimally thin Debye layer. 
Each effect is governed by different length scales -- for instance, the finite Debye layer effect is determined by the local radius of curvature $r_c$ and the Debye length $\kappa^{-1}$, whereas the confinement effect is set by the gap spacing between the ellipsoid and the wall, which is sensitive to the semiminor axis of the ellipsoid $a_2$ and the channel half height $h$.
Therefore, the relevant parameters that characterize the curvature effect $\lambda_1$ and the confinement effect $\lambda_2$ are $\lambda_1=\lambda_1(\kappa^{-1},r_c)$ and $\lambda_2=\lambda_2(h,a_2)$. 

A prolate ellipsoid oriented parallel to the gradient will experience interfacial flow occurring axisymmetrically along the meridian, whereas the flow will mostly take place along the geodesics across the minor axis plane for ellipsoids oriented transverse to the gradients (akin to long transverse cylinders). This implies that the relevant radius of curvature determining the mobility will vary depending on the orientation. For longitudinal ellipsoids, we may choose the characteristic radius of curvature to be the area-averaged principal radius of curvature along the meridian $r_{c_1}$,  i.e., $r_c^\parallel=\int_S r_{c_1} dA/ \int_S dA$, where 
\begin{equation}
	r_{c_1}=\frac{\left[(a_2 \mathrm{cos}\,\varphi)^2+(a_1 \mathrm{sin} \,\varphi)^2\right]^{3/2}}{a_1a_2},
\end{equation}
where $\varphi$ is the polar angle. 
On the other hand, 
the characteristic radius of curvature for transverse ellipsoids is effectively the semiminor axis, i.e., $r_c^{\perp}=a_2$.
Using these orientation-dependent characteristic radii of curvature $r_c^{(\parallel, \perp)}$, we attempt to approximate the ellipsoid as a sphere whose effective radius is $r_c^{(\parallel, \perp)}$. 
This assumption is based on the observations from previous literature, where the phoretic mobilities of cylinders and spheroids share the same functional forms as the spheres, and those mobilities become similar with decreasing Debye layer thickness \cite{obrien1988electrophoresis,yoon1989electrophoresis,dukhin1993non,keh1993diffusiophoresis}. Notably, the mobility of a transverse cylinder is remarkably close to that of the sphere of the same radius \cite{dukhin1993non,keh1993diffusiophoresiscylinder,stigter1978electrophoresis,ohshima1996henry,ohshima2021approximate}, which also confirms that the local radius of curvature \textit{along} the flow direction is the relevant length scale for the curvature effect. 

In this spirit, we may further approximate the ellipsoid as a sphere, which will allow making use of the well-characterized theories for spheres whose analytical expressions are readily available, instead of taking a numerical approach to compute the mobility for ellipsoids. 
Then, $M_\infty$, $\Phi_1$, and $\Phi_2$ can be expressed using analytical models for charged spheres immersed in 1:1 electrolytes as \cite{Prieve1984,chen2005diffusiophoresis} 
\begin{eqnarray}
		M_\infty = (\epsilon/\eta)(kT/e) [(\beta\widetilde{\zeta} ) + 4 \mathrm{ln\,cosh}(\widetilde{\zeta} / 4) ],\\
		\Phi_1(\lambda_1=\kappa^{-1}/r_c)= 1/(1-\alpha\lambda_1),\\
			\Phi_2(\lambda_2=a_2/h)= 1-\gamma \lambda_2^3 + \delta \lambda_2^5,
\end{eqnarray}
where $\epsilon$ is the fluid permittivity, $\eta$ is the fluid viscosity, $\widetilde{\zeta}=\zeta e/kT$ is the dimensionless zeta potential, and $\beta=(D_+-D_-)/(D_++D_-)$ is the dimensionless solute diffusivity contrast, where $D_\pm$ is the ion diffusivity. 
Here, $\alpha$ is a series of integrals of exponential functions, where the exact formulation can be found in \cite{Prieve1984},
and $\gamma\approx0.27$ and $\delta\approx0.34$ are factors for a charged sphere undergoing diffusiophoresis along the midplane of a planar slit of gap height $2h$ \cite{chen2005diffusiophoresis}. 
As mentioned, we choose $r_c=r_c^\parallel$ for estimating the longitudinal mobility  ($M^\parallel$), whereas $r_c=a_2$ for transverse mobility  ($M^\perp$). 

The ellipsoid mobility approximated as a sphere under confinement using Eqs.~(3)--(5) is shown in Fig.~4b (blue curve). 
The combined use of idealized theories for spheres effectively captures the key experimental observations on the nonmonotonic mobility under confinement without involving any free parameters. 
We can also describe the orientation-dependent diffusiophoresis by taking the $x$-direction components of $M^\parallel$ and $M^\perp$  of an ellipsoid oriented at an angle $\theta$, such that 
\begin{equation}
	M(\theta)=\left[(M^\parallel \mathrm{cos} \,\theta)^2 + (M^\perp \mathrm{sin} \,\theta)^2\right]^{1/2}.
\end{equation}
Likewise, the orientation-dependent mobility is also well characterized by the sphere approximation, shown as the green curve in Fig.~2c.

We note that the Brownian translational diffusion, which is also anisotropic for nonspherical particles \cite{happel2012low}, has a negligible contribution to the observed orientation-dependent motion.  
The anisotropic diffusivity of ellipsoids in free space is given as $D^{(\parallel,\perp)}=kT/f^{( \parallel,\perp)}$, where $f^\perp=8\pi\eta a_1/[\mathrm{ln}(2a_1/a_2)+1/2] \approx 2f^\parallel$
are the transverse and longitudinal drag coefficients of slender prolate ellipsoids (AR~$\gg$~1) \cite{tillett1970axial}.
For the ellipsoids used in Fig.~2 (AR~=~6.5), the particle diffusivity should be much smaller than the free-space diffusivity $D^{(\parallel,\perp)} \approx 10^{-13}$~m$^2$/s due to the additional boundary confinement \cite{han2009quasi}.
Since the diffusiophoretic mobility $M = \mathcal{O}(10^{-10}$~m$^2$/s) is at least three orders of magnitude larger than the translational diffusivity without the confinement, we can effectively neglect the orientation-dependent Brownian translation.

To summarize, we have demonstrated experimentally that the diffusiophoresis of colloidal ellipsoids can be sensitive to their orientation as well as shape. The experimental conditions of our system, where the particles are microscale and the surroundings provide physical confinement, led to peculiar colloid behaviors, such as the lack of preferential orientation and the non-monotonic mobility. 
As such systems represent environments often found in biological, biomedical, and geological systems, our results may provide insights into a number of important transport problems that involve microscale slender objects under confinement, such as bacteria swimming \cite{bhattacharjee2019bacterial,lynch2022transitioning} or cell migration \cite{friedl2011cancer,stroka2014water} in complex microenvironments, nanomedicine delivery in compressed tissues \cite{jain2010delivering,doan2021confinement}, and contaminant transport in the subsurface \cite{syngouna2013cotransport}.  

\begin{acknowledgments} 
This material is based upon work supported by the National Science Foundation under Grants No. 2200882 and 2223737. S.S. thanks Jesse Ault for valuable discussions.
Y.S. thanks the National Science Foundation under Grants No. 1705745 and No. 2300317.
\end{acknowledgments}

\bibliography{reference}

\end{document}